\documentclass[twocolumn,10pt]{article} 

\usepackage[square,numbers,sort&compress,comma]{natbib}

\usepackage{amsmath}
\usepackage{amssymb}
\usepackage{caption}
\usepackage{graphicx}
\usepackage{latexsym}
\usepackage{times}
\usepackage[pagewise]{lineno}
\usepackage{hyperref}
\usepackage{algorithm}
\usepackage{algorithmic,color}
\usepackage[version=4]{mhchem}

\topmargin - 12pt 
\renewenvironment{abstract}%
              {
               \small
               {\bfseries \abstractname}
               \par
               \vspace{10pt}
              }

\renewcommand\abstractname{Abstract}

\newcommand{\nomenclature}
              [1]
              {
               \bgroup
               \flushleft
               \small\bf
               #1
               \par
               \egroup
              }

\renewcommand{\section}
              [1]
              {
               \bgroup
               \flushleft
               \small\bf
               \refstepcounter{section}
               \arabic{section}. #1
               \par
               \egroup
              }

\renewcommand{\subsection}
              [1]
              {
               \bgroup
               \flushleft
               \small\em
               \refstepcounter{subsection}
               \arabic{section}.
               \arabic{subsection}. #1
               \par
               \egroup
              }

\renewcommand{\subsubsection}
              [1]
              {
               \bgroup
               \flushleft
               \small\em
               \refstepcounter{subsubsection}
               \arabic{section}.
               \arabic{subsection}.
               \arabic{subsubsection}. #1
               \par
               \egroup
              }

  \newcommand{\acknowledgement}
              [1]
              {
               \bgroup
               \flushleft
               \small\bf
               #1
               \par
               \egroup
              }

  \newcommand{\sectionbib}
              [1]
              {
               \bgroup
               \flushleft
               \small\bf
               #1
               \par
               \egroup
              }

\begin{document}

\title{\LARGE A comprehensive study on the accuracy and generalization of deep learning-generated chemical ODE integrators}

\author{{\large Han Li$^{a,b,1}$, Ruixin Yang$^{a,b,1}$, Yangchen Xu$^{a}$, Min Zhang$^{a,b}$,}\\[10pt]{\large Runze Mao$^{a,b}$, Zhi X. Chen$^{a,b,*}$}\\[10pt]
        {\footnotesize \em $^a$State Key Laboratory of Turbulence and Complex Systems, Aeronautics and Astronautics, College of Engineering,}\\[-5pt]
        {\footnotesize \em Peking University, Beijing, 100871, China}\\[-5pt]
        {\footnotesize \em $^b$AI for Science Institute (AISI), Beijing, 100080, China}\\[-5pt]}

\date{}


\small
\baselineskip 10pt


\twocolumn[\begin{@twocolumnfalse}
\vspace{50pt}
\maketitle
\vspace{40pt}
\rule{\textwidth}{0.5pt}
\begin{abstract} 
The application of deep neural networks (DNNs) holds considerable promise as a substitute for the direct integration of chemical source terms in combustion simulations. However, challenges persist in ensuring high precision and generalisation across various different fuels and flow conditions. In this study, we propose and validate a consistent DNN approach for chemistry integration in a range of fuels and premixed flame configurations. This approach generates thermochemical base state from a set of low-dimensional laminar flames, followed by an effective perturbation strategy to enhance the coverage of the composition space for higher generalisation ability. A constraint criterion based on heat release rate is then employed to remove the nonphysical perturbed states for improved accuracy. 
Without specific tuning, three DNNs are consistently trained for three representative fuels, i.e., hydrogen, ethylene and Jet-A. Comprehensive validations are conducted using 1-D laminar flames and two typical turbulent premixed flames. The DNN model predictions on various physical characteristics, including laminar and turbulent flame speeds, dynamic flame structures influenced by turbulence-chemistry interactions, and conditional scalar profiles, all exhibit good agreement with the results obtained from direct integration. This demonstrates the exceptional accuracy and generalisation ability of the proposed DNN approach. Furthermore, when the DNN is used in the simulation, a significant speed-up for the chemistry integration is achieved, approximately 50 for the ethylene/air flame and 90 for the Jet-A/air flame.
\end{abstract}
\vspace{10pt}
\parbox{1.0\textwidth}{\footnotesize {\em Keywords:} Machine learning; Chemistry integration; Turbulent combustion modelling; Deep Neutral Network}
\rule{\textwidth}{0.5pt}
\vspace{10pt}

*Corresponding author.\\
\textit{E-mail address:} chenzhi@pku.edu.cn (Zhi X. Chen).

\end{@twocolumnfalse}] 

\clearpage

\section{Introduction\label{sec:section1}} \addvspace{10pt}

The utilization of finite rate chemistry (FRC) in combustion modeling typically yields a more comprehensive and accurate depiction of reaction processes and flame dynamics.
However, detailed FRC modeling entails a substantial computational cost driven by direct integration (DI) of stiff ordinary differential equations (ODEs)~\cite{poinsot2005theoretical}. To improve computational efficiency, various methods for chemical mechanism reduction~\cite{lam1994csp,Sun10,Lu05,Pepiot08}
have been introduced. 
However, even with a reduced mechanism, which typically retains several tens of chemical species exhibiting markedly distinct chemical time-scales, DI of the associated ODEs remains a expensive task, constituting over 80\% of the total computational cost in practical combustion simulations. 

To tackle the challenge of balancing accuracy and efficiency, recent strides in artificial intelligence for scientific applications, notably in ML, have opened up innovative perspectives~\cite{ihme2022combustion}. In the present work, we focus on employing ML models to replace DI with a similar level of accuracy. As an pioneering work, Christo et al.~\cite{christo1995utilising} developed a neural network (NN) to represent a three-step mechanism in the joint PDF simulation of turbulent jet flames, showcasing the substantial potential of ML in combustion modelling. Blasco et al.~\cite{blasco1998modelling,blasco2000self} proposed a self-organizing map (SOM) approach with an enhanced precision in predicting the temporal evolution of reactive species.

However, the aforementioned works relied on training the NN on the specific problem of interest with limited range of applicability. To address this issue, Sen et al. ~\cite{sen2009turbulent,sen2010linear} obtained thermochemical states from direct numerical simulation (DNS) and Linear Eddy Mixing (LEM) model calculations, and proved successful in simulating syngas/air flames. Chatzopoulos et al.~\cite{chatzopoulos2013chemistry} collected samples from non-premixed laminar flames and using the SOM technique to train NN models. These models were subsequently applied to Reynolds Averaged Navier–Stokes (RANS)-PDF simulations of DLR jet flames. In a follow-up work, Franke et al.~\cite{franke2017tabulation} integrated extinguishing flamelets into the training dataset. Recently, Wan et al.~\cite{wan2020chemistry} generated training samples from a non-premixed micro-mixing canonical problem and then achieved good agreement with the DI approach in a syngas turbulent oxy-flame. Ding et al.~\cite{ding2021machine} and Readshaw et al.~\cite{readshaw2021modeling} collected samples from numerous 1-D laminar flames and randomised these data for training using multiple NNs (three for each species). The resulting NNs were tested to be effective on methane/air flames, including one-dimensional laminar flame and Sandia turbulent jet flames. In contrast to sampling from low-dimensional flames, Zhang et al.~\cite{zhang2022multi} introduced a multi-scale sampling method to collect data from the full composition space. Owing to the powerful fitting capability of deep neural network (DNN), They trained a rather \textit{big model} using a large dataset comprising over 5 million samples, covering a broad composition space for hydrogen/air flames. With over 1.6 million model parameters, this DNN showed a good generalisation capability across a range of laminar and turbulent flames under various conditions.

The usefulness of ML models essentially relies on achieving both high accuracy and generalisation ability. Previous studies employing a multiple layer perceptron (MLP) architecture such as the SOM-MLP approach~\cite{blasco1998modelling,chatzopoulos2013chemistry} and MMLP~\cite{readshaw2021modeling} have demonstrated improved accuracy. Effective generalisation has also been achieved by collecting training data from simple canonical problems~\cite{chatzopoulos2013chemistry,franke2017tabulation,wan2020chemistry,ding2021machine,readshaw2021modeling}. However, the validation of these ML models was predominantly limited to canonical problems and a specific multi-dimensional turbulent flame, leaving the generalization for other turbulent flame configurations unexplored. Furthermore, prior research has primarily focused on a single and simple chemical system with a small number of chemical species, such as hydrogen, methane, and syngas. Limited attention has been given to the generalization ability across different and complex fuels, like large hydrocarbons, whose reduced mechanisms comprise dozens of chemical species. 

With this motivation, the primary objective of the present work is to develop a consistent and robust methodology for generating generic samples, training high-precision DNNs, and comprehensively assessing their validity across a spectrum of fuels ranging from simple hydrogen to complex kerosene and in different turbulent premixed flame configurations. 
Three DNNs are trained for reactive mixtures of hydrogen/air with a 9-species mechanism~\cite{burke2012comprehensive}, ethylene/air with a 24-species mechanism~\cite{wu2017develop} and Jet-A/air with a 41-species mechanism~\cite{xu2018physics}, respectively. The model deployment and {\em a posteriori} assessment are then performed using our recently developed open-source code DeepFlame~\cite{Mao2023DeepFlame}, which interfaces OpenFOAM, Cantera and PyTorch libraries. 
Two typical turbulent premixed flame cases are considered: a temporally evolving jet flame and a propagating flame kernel in homogeneous isotropic turbulence (HIT). All DNN models, CFD codes and test cases presented in this study are made available for community data sharing and reproducibility\footnote{\href{https://github.com/deepmodeling/deepflame-dev}{Available at: github.com/deepmodeling/deepflame-dev}}.

The remainder of this paper is organised as follows. Section 2 discusses the methodology for generic DNN training. In Section 3, we present the turbulent case setups for model validation. In Section 4, the results for different fuels and chemical mechanisms are discussed. Conclusions are summarised in Section 5.

\section{Deep learning methodology\label{sec:section2}} \addvspace{10pt}

This section describes the step-by-step procedures for the proposed DNN approach including training data generation and sampling, network design and learning, and model prediction test. These are all kept consistent for a range of fuels and turbulent flame configurations considered in this study. 

\subsection{Thermochemical base state generation\label{subsec:subsection1}} \addvspace{10pt}

Due to the highly non-linear and stiff nature of chemical ODE systems, the error tolerance allowed for reaction rate integrator is extremely stringent. Thus, it is imperative to generate representative training data that encompasses a proper distribution of thermochemical states and thoroughly covers the relevant composition space. 

Instead of directly exploring the high-dimensional thermochemical sampling space,  
we first locate a low-dimensional manifold region as the base states. This does not necessarily imply a flamelet assumption but provides a good start point for the sampling process. 
In this study, the thermochemical base states are collected from simulations of a set of canonical laminar premixed flames to ensure generalisation and also to minimise the complexity of data generation. This approach can be easily extended to diffusion flames and other canonical configurations. 
The computational domain of these 1-D laminar flames is initialised with premixed fuel/air mixture in one half and equilibrium states are set in the other half. The initial conditions (i.e., temperature, pressure and equivalence ratio) are set according to the global parameters of the target turbulent flames. Simulations are conducted until steady states are reached. The simulated time of each 1-D flame is estimated to be around ten times the respective chemical time scale $\tau_{chem}=\delta_L/S_L$, and the temporal thermochemical states are sampled every 100 simulation time steps.

\subsection{Data perturbation and augmentation\label{subsec:subsection2}} \addvspace{10pt}

The thermochemical states obtained from laminar flames might not comprehensively cover the relevant composition space in {\em a posteriori} applications. More critically, these particular states follow an exact path in the sample space (essentially flamelet manifolds) and hence the trained model is susceptible to perturbations, i.e. deviation from the manifold in the thermochemical state. 
To address this issue and enhance model robustness, a data augmentation strategy is applied to perturb the collected states, mimicking the multi-dimensional transport and turbulence perturbations. At each sample point, temperature, pressure and inert species are randomly perturbed using~\cite{readshaw2021modeling}
\begin{equation}
    x_{R} = x + \alpha*\beta*(x_{max} - x_{min}), 
\end{equation}
where $x_{R}$ and $x$ represent the temperature, pressure or inert species mass fraction of the perturbed sample and the original sample, respectively. The perturbation amplitude $\alpha$ is user-specified and $\beta$ is a uniformly randomised number within the range (-1,1). Given the significant changes in mass fraction magnitudes for reactive chemical species, a different exponential randomisation strategy is implemented: 
\begin{equation}
    y_{R} = y^{1+\alpha*\beta},
\end{equation}
where $y_{R}$ and $y$ represent the species mass fraction (excluding inert species) of the perturbed sample and the original sample, respectively. The resulting randomly generated mass fractions of these species are normalised to $1-x^{N_2}_{R}$ to ensure mass conservation.

\begin{figure}[t]
\centering
\includegraphics[width=180pt]{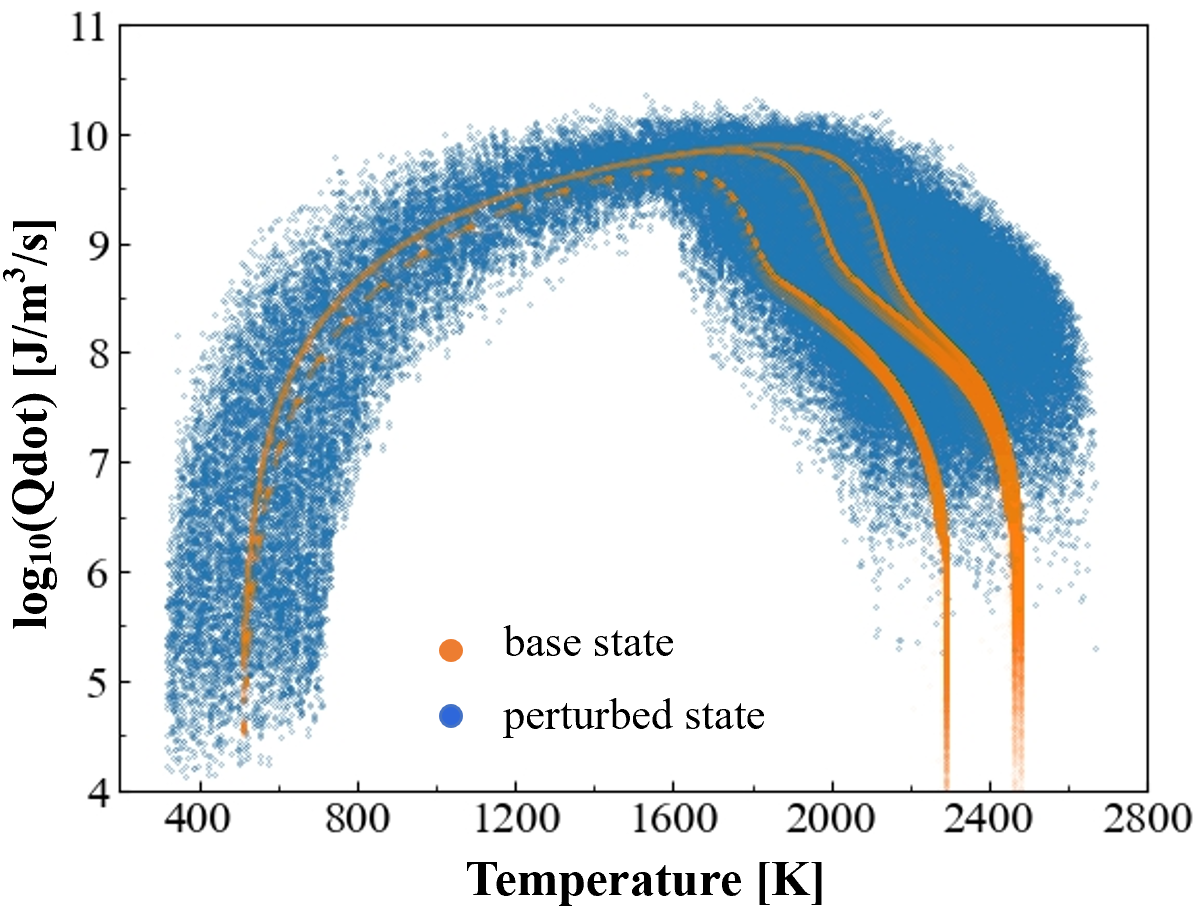}
\caption{\footnotesize Joint distribution of heat release rates and temperature of ethylene/air thermochemical states.}
\label{Qdot_T}
\end{figure}

The perturbation amplitude $\alpha$ and number of randomisation times $N_R$ can be adjusted according to the amount of collected states and the turbulence intensity of the flow field. In this work, we use $\alpha \in [0.1,0.15]$ and $N_R=10$ perturbations for all three fuels considered. This practice substantially enhances the coverage of the composition space. However, it may generate numerous nonphysical states if left unconstrained, which effectively lowers the model accuracy in the regions of interest. To address this, a threshold criterion, based on the heat release rate change between the original collected state and the perturbed state, is used to remove the nonphysical perturbed states for improved accuracy and generalization ability. The sample space distribution before and after the data augmentation is illustrated in Fig.~\ref{Qdot_T}, where the manifold-like orange symbols represent states collected from 1-D laminar flames, and the scattered blue symbols depict the physics constrained random perturbation states. 


\begin{figure*}[t]
\centering
\includegraphics[width=360pt]{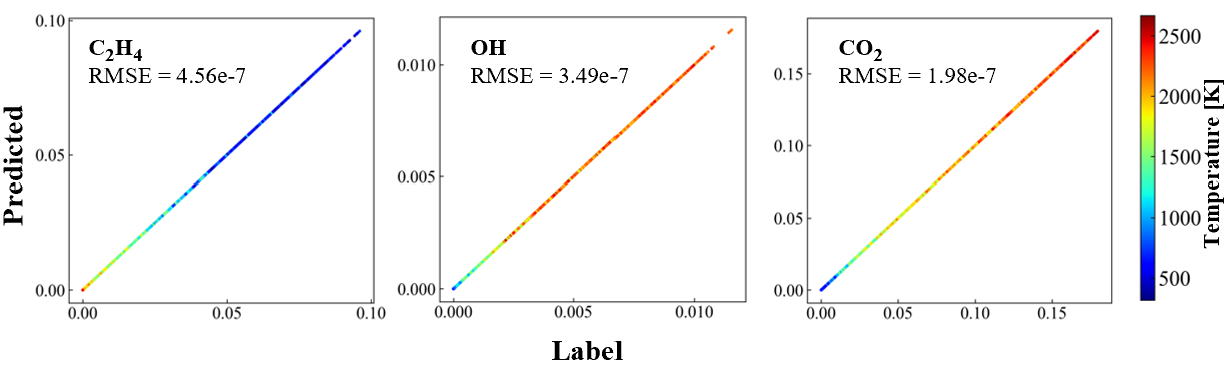}
\\
\vspace{8 pt}
\caption{\footnotesize DNN prediction for typical major and minor species mass fractions.}
\label{single_predict}
\end{figure*}

\begin{figure}[ht]
\centering
\includegraphics[width=180pt]{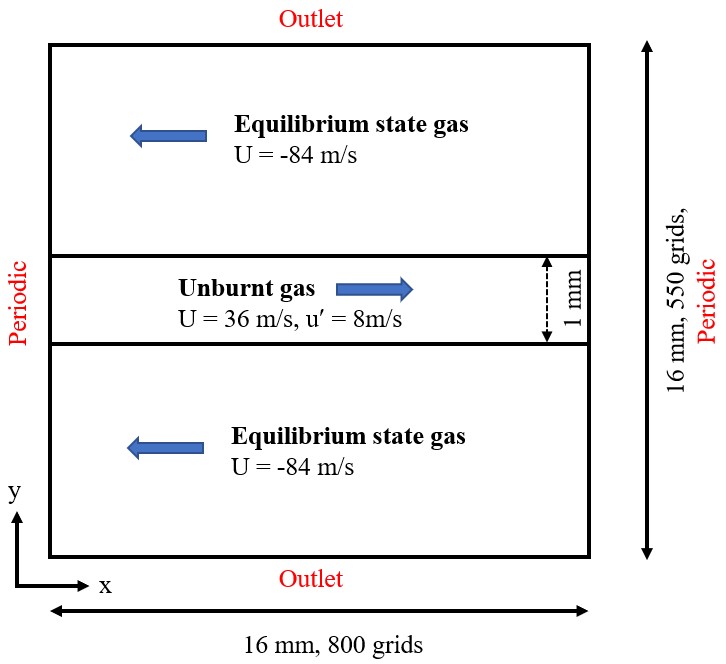}
\\
\vspace{8 pt}
\caption{\footnotesize Computational domain of the 2-D evolving jet flame.}
\label{2Devolving_domain}
\end{figure}

\subsection{Deep neural network\label{subsec:subsection3}} \addvspace{10pt}

The DNN input layer includes temperature, pressure and species mass fractions, represented as 
\begin{equation}
    \boldsymbol{x}(t)=\{T(t), P(t), \mathcal{F}[\boldsymbol{Y}(t)]\}.
\end{equation}
The output layer consists of the change of species mass fraction over a given time step size, denoted as 
\begin{equation}
    \boldsymbol{u}^*[\boldsymbol{x}(t);\Delta t]=\mathcal{F}[\boldsymbol{Y}(t+\Delta t)] - \mathcal{F}[\boldsymbol{Y}(t)], 
\end{equation}
where $\mathcal{F}$ is the Box-Cox transformation (BCT) ~\cite{box1964analysis} employed for the multi-scale species mass fractions. This transformation provides a more uniform distribution of small-scale clustered sample data, thereby enhancing the performance of the neural network in predicting the species mass fraction changes. 
The training dataset $D=\{\boldsymbol{x_i},\boldsymbol{u^*_i}\}^N_{i=1}$ undergoes $Z$-score normalisation, involving the subtraction of the mean and division by the standard deviation, for both input $\boldsymbol{x_i}$ and output $\boldsymbol{u^*_i}$. Here, $N$ represents the sample size. To ensure accuracy in predictions across the output, each species mass fraction is trained and predicted individually; however, the DNNs are assembled in one integrated model after training. Each DNN consists of three hidden layers with 1600, 800 and 400 perceptrons. The activation function used for the network is the Gaussian Error Linear Unit (GELU), and hyper-parameter optimisation is performed using the Adam algorithm. It is important to note that the DNN predictions yield only the mass fractions. The temperature and density are subsequently computed based on enthalpy and the mass conservation laws.

A common loss function for the DNN output constrain is given by
\begin{equation}
\begin{split}
    \mathcal{L}&=\frac{1}{N} \sum_{i=1}^N \left \vert \boldsymbol{u}^*_i - \boldsymbol{u}_i \right \vert \:,
\end{split}
\label{Loss}
\end{equation}
where $\boldsymbol {u}_i$ is the DNN output. In this study, we incorporate three novel additional principles into the loss function, including mass fraction unity conservation, energy conservation, and considerations related to heat release rate, which substantially improves the model accuracy. Generally, the training $||L1||$ loss on $\mathcal{F}[\boldsymbol{Y}(t)]$ is of the order of $10^{-4}$, and more details can be found in the shared code included in the Supplementary Material.

\subsection{A priori assessment\label{subsec:subsection4}} \addvspace{10pt}

As a standard procedure in deep learning, once the DNN model is trained {\em a priori} test is performed to evaluate the prediction errors. This step is crucial before subsequent validation using reacting flow cases. As an example, here we consider the DNN model for ethylene/air mixture and the results are similar for other fuels. Figure~\ref{single_predict} shows the predictions for the fuel species $\ce {C2H4}$, an important radical $\ce {OH}$ and a product species $\ce {CO2}$. It can be seen that the predicted values are in excellent agreement with the randomly chosen label values. The root mean square errors (RMSE) are of order of $10^{-7}$, which is expected to satisfy the chemical ODE requirement in various laminar and turbulent flames.


\begin{figure*}[t]
\centering
\includegraphics[width=300 pt]{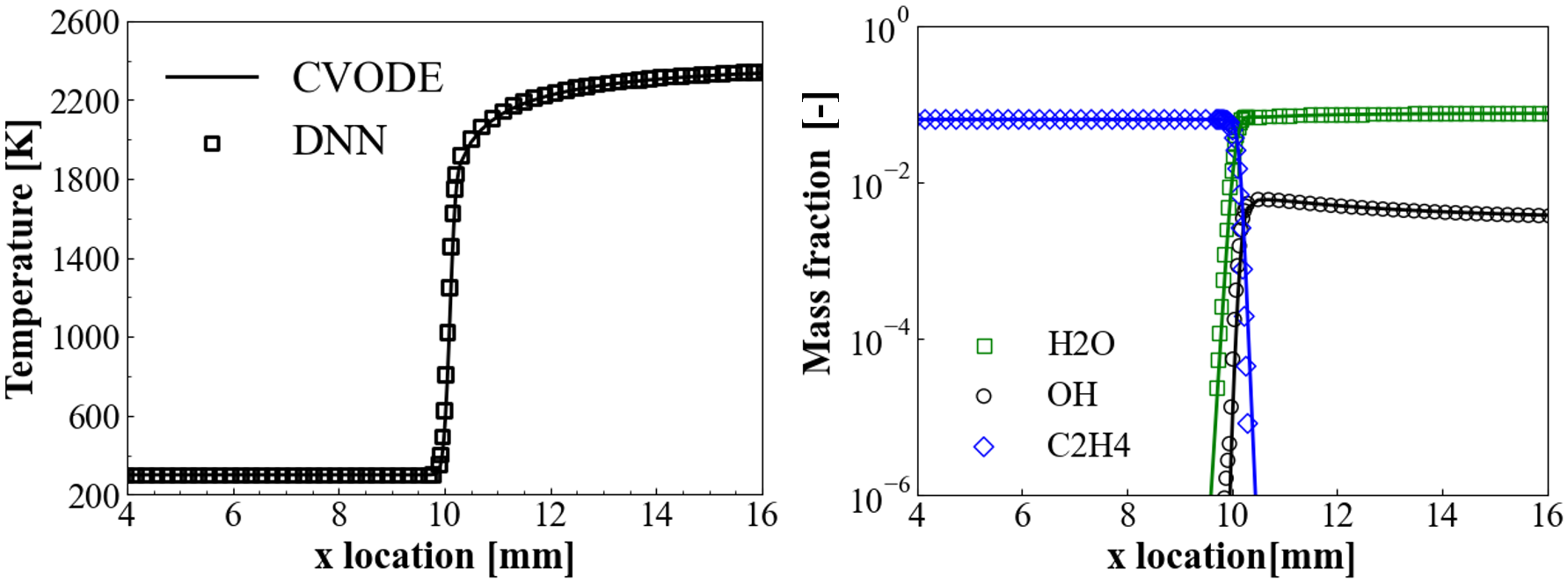}
\\
\vspace{8 pt}
\caption{\footnotesize Spatial distribution of temperature (left) and major species (right) at P = 1 atm, $\phi$ = 1.0 and $T_u$ = 300 K for ethylene/air.}
\label{Temp_Y_distri}
\end{figure*}

\begin{figure}[ht]
\centering
\includegraphics[width=140 pt]{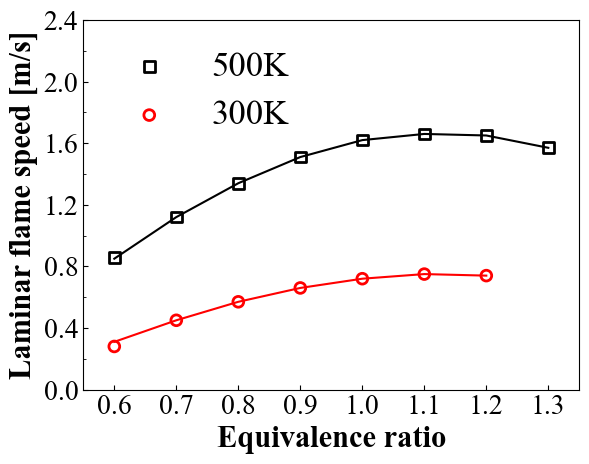}
\\
\vspace{8 pt}
\caption{\footnotesize Laminar flame speed at different equivalence ratios for ethylene/air mixture.}
\label{SL_phi}
\end{figure}

\section{Test case setup\label{sec:section3}} \addvspace{10pt}

To validate the trained DNN models in combustion simulations, two 2-D premixed turbulence flame cases are designed and described in this section.

\subsection{Temporally evolving turbulent jet flame\label{subsec:subsection1}} \addvspace{10pt}

Turbulent planar jets are prototypical free shear flows, which are widely used to study turbulence-chemistry interaction~\cite{driscoll2020premixed}. Here, we present a two-dimensional temporally evolving planar turbulent jet flame, considering the mixing and reaction processes of scalars in turbulent shear flows. A similar configuration has also been utilized by Satio et al.~\cite{saito2023data} to validate their DNN model for ammonia combustion.

As shown in Fig.~\ref{2Devolving_domain}, a square computational domain of $L=16$ mm is considered, initially filled with stoichiometric fuel/air mixture in the central region and an equilibrium state gas elsewhere. To initialise turbulent shear flow, the internal velocity field is generated from a precursor non-reactive jet flow simulation using the synthetic eddy turbulence inflow generator and then superimposed onto the unburnt gas region. Periodic boundary conditions are applied on the left and right sides, while outlet conditions are set for the top and bottom boundaries. The domain is discretised with $800\times550$ grids, with a minimum grid size of 20$\mu m$ to ensure proper resolution on the flame front. The grid is uniform in the $x$-direction and stretched at both ends in the $y$-direction. 


\subsection{Ignition in homogeneous isotropic turbulence\label{subsec:subsection2}} \addvspace{10pt}

The second test case involves a flame kernel ignition of premixed mixture in two-dimensional homogeneous isotropic turbulence (HIT). This simulation setup features an ignition to propagation transition process, highlighting the turbulence effect on the flame evolution~\cite{Karimkashi2020a}. Therefore, it serves as a challenging validation case for the DNN models.

In the simulation, a square computational domain of $L\times L=10\pi \times 10\pi$ mm$^2$ is used, initialised with premixed stoichiometric fuel/air mixture at a given temperature and pressure. To ignite the mixture, a circular hot spot with a radius of $L/10$ filled with equilibrium gases is placed in the center of the domain. The HIT generation approach in~\cite{Vuorinen2016DNSLab} is adopted and the fully evolved velocity field is then mapped to the computational domain as the initial flow field. Boundary conditions are set to zero gradient for temperature and species mass fractions, and a non-reflective wave transmissive condition is used for pressure and velocity. The domain is uniformly discretised with 1024 × 1024 grids to ensure good resolution for both the flame and turbulence. The simulations are continued until the full domain is ignited.

\begin{figure*}[ht]
\centering
\includegraphics[width=320pt]{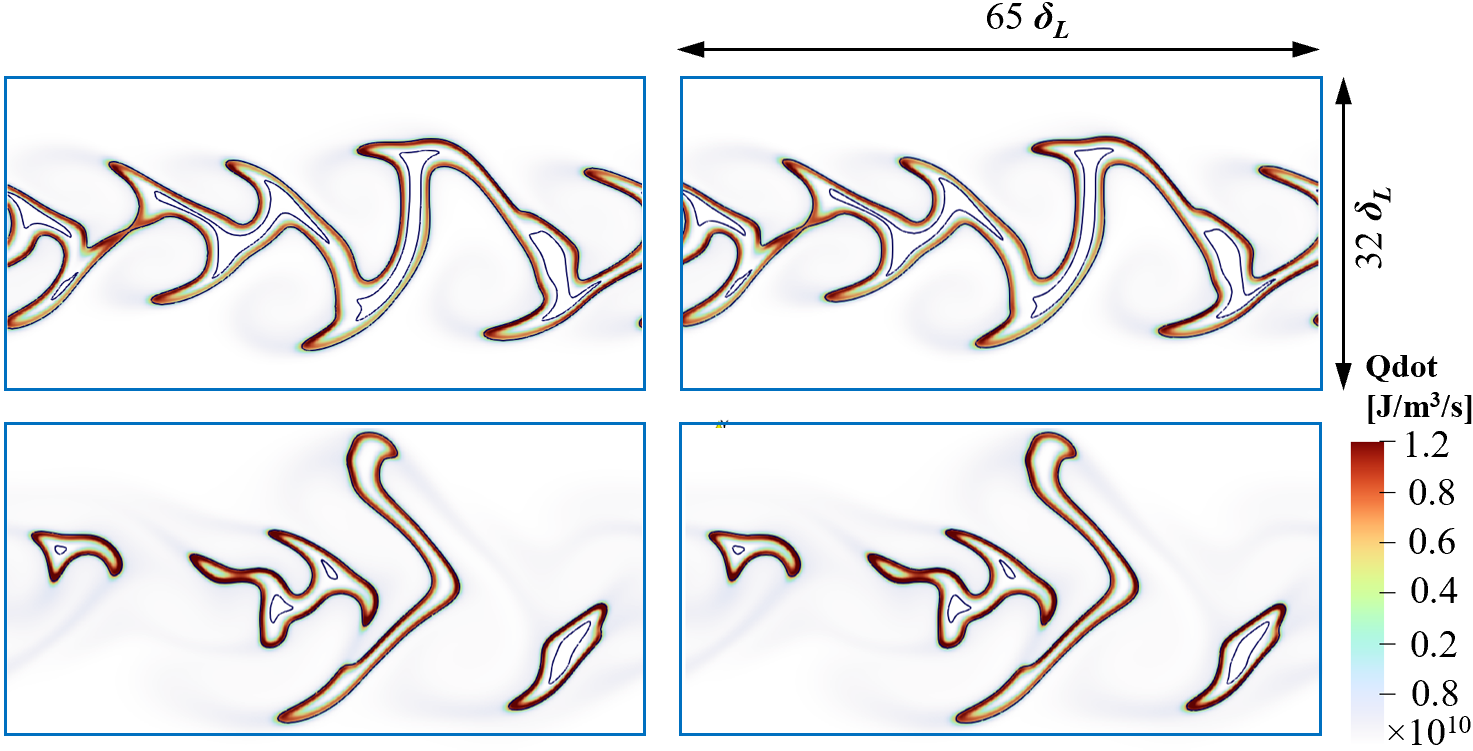}
\\
\vspace{8 pt}
\caption{\footnotesize Contours of heat release rate and isolines (black lines) for \ce{CH2O} with a mass fraction of 5e-4 predicted using CVODE (left) and the DNN model (right) in the evolving jet flame of ethylene/air mixture at t = 0.12, 0.16 ms.}
\label{C2H4_evolving_Qdot}
\end{figure*}

\begin{figure*}[ht]
\centering
\includegraphics[width=390pt]{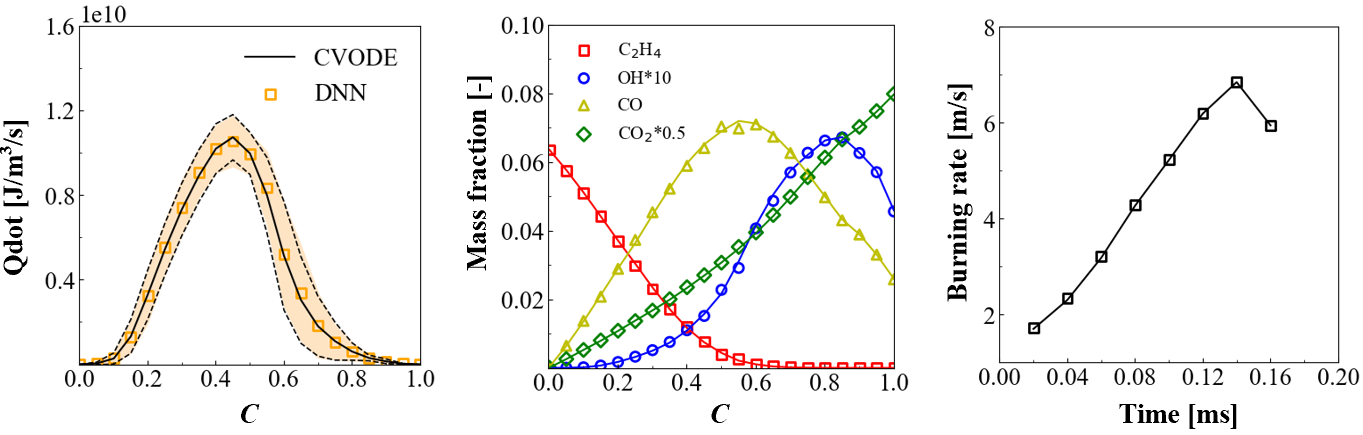}
\\
\vspace{8 pt}
\caption{\footnotesize Conditionally averaged Qdot (left, with standard deviation plotted) and mass fraction (middle) at  t = 0.12 ms, and turbulent burning rate (right) in the evolving jet flame for ethylene/air mixture (lines from CVODE, symbols from DNN).}
\label{C2H4_conditional_mean}
\end{figure*}

\section{Results and Discussion\label{sec:section4}} \addvspace{10pt}

This section presents extensive validations of the proposed methodology and DNN models in various 1-D laminar flames and 2-D turbulent flames coupling the effects of convection, stretching and turbulence. 
All cases are scale-resolved using detailed numerical simulation with detailed chemistry and mixture-averaged transport~\cite{Mao2023DeepFlame}. 
The predictions using DNN models are validated against the results obtained through DI using the Cantera CVODE solver. For conciseness, results for the relatively simple hydrogen/air combustion are provided in the Supplementary Material.

For ethylene/air combustion, thermochemical states are collected from three laminar flame cases with an unburnt gas temperature $T_u$ of 500 K and pressure of 1 atm. The equivalence ratios for the three cases are 0.8, 1.0 and 1.2, respectively. Subsequently, approximately 200,000 state points are collected, perturbed and augmented to generate a training set of around 800,000 samples. A time-step size of $10^{-7}$~s is used for DNN model prediction. The training process typically evolves for 2000 epochs with an initial learning rate of 0.001, which decreases 10 times every 200 epochs. For Jet-A combustion, the unburnt gas temperature is specified as 800 K, while keeping other conditions consistent with the ethylene case. Next, the resulting DNN models are tested comprehensively in 1-D laminar flames and 2-D turbulent flames.

\subsection{1-D premixed laminar flame\label{subsec:subsection1}} \addvspace{10pt}

In Fig.~\ref{Temp_Y_distri}, the spatial distributions of temperature and major species mass fraction are presented for the ethylene/air laminar flames at $T_u= 300$~K. The flame front position and flame structure predicted by the DNN model align closely with the results obtained through direct integration using CVODE. This demonstrates the high precision of the trained DNN models in predicting chemical kinetics in premixed laminar flames. Furthermore, Fig.~\ref{SL_phi} presents the laminar flame speeds for ethylene/air mixture at various equivalent ratios ranging from 0.6 to 1.3, extending beyond the sampling range (0.8 to 1.2). The predictions by the DNN model closely match those from CVODE, demonstrating an excellent robustness of the data perturbation and augmentation approach which significantly enhances the generalization and even extrapolation capability of the DNN model.



\subsection{2-D evolving jet flame\label{subsec:subsection2}} \addvspace{10pt}

Figure~\ref{C2H4_evolving_Qdot} depicts the contours of heat release rate in the evolving jet flame for stoichiometric ethylene/air mixture. Iso-lines of key intermediate \ce{CH2O} at a mass fraction of 5e-4 are also shown to assess fine radial structure prediction. Again, an excellent agreement is observed between the DNN and DI results for both time instants considered. For a more quantitative comparison, conditional heat release rate and mass fractions in the progress variable space are plotted in Fig.~\ref{C2H4_conditional_mean}. The progress variable is defined using a combined mass fraction of \ce{H2O} and \ce{CO2} normalised by their burnt values. It can be seen that all the conditional profiles given by the DNN agree well with the DI results, suggesting a high accuracy under turbulent flame conditions. 

In addition, the turbulent burning velocity is calculated using 
\begin{equation}
    S_{T}=\frac{1}{A} \int_{V} \frac{\dot{\omega}_{T}}{c_{p}\left(T_{b}-T_{u}\right)} d V \:,
\end{equation}
where $\dot{\omega}_{T}$ is the heat release rate, $c_{p}$ is the specific heat capacity at constant pressure, $T_{b}$ and $T_{u}$ are the burnt and unburnt gas temperature respectively, and $A$ is the equivalent flame front area. The temporal evolution of $S_{T}$ in Fig.~\ref{C2H4_conditional_mean} shows no observable difference between the predictions of the DNN model and CVODE, further confirming the exceptional precision of the DNN model.




\begin{figure}[ht]
\centering
\includegraphics[width=160pt]{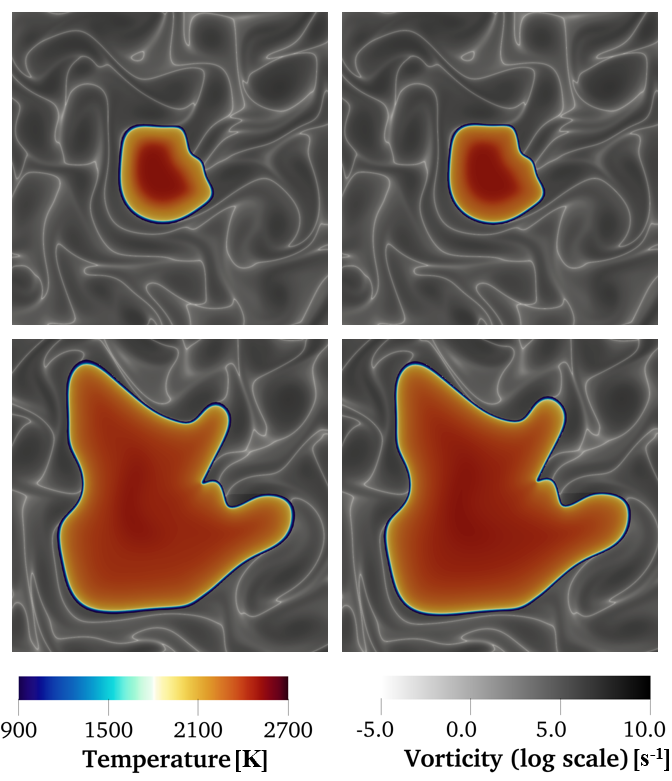}
\\
\vspace{8 pt}
\caption{\footnotesize Contours of flame temperature and flow vorticity predicted using CVODE (left) and the DNN model (right) at $t = 0.4, 1.2$ ms in the HIT flames of JetA/air mixture.}
\label{Jet_HIT_T_Vor}
\end{figure}

\begin{figure}[ht]
\centering
\includegraphics[width=140 pt]{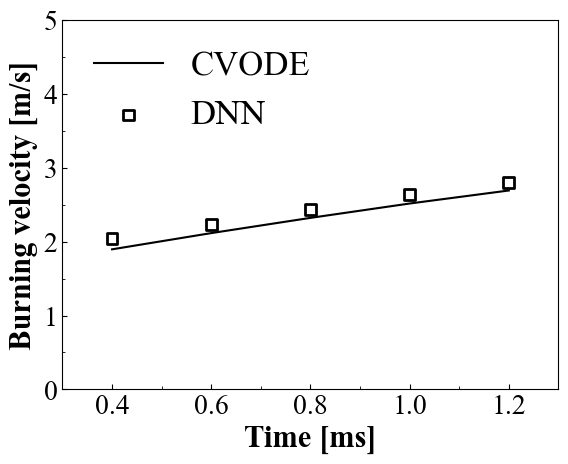}
\\
\vspace{8 pt}
\caption{\footnotesize Comparisons of turbulent burning velocity in the HIT flames of JetA/air mixture.}
\label{Jet_HIT_St_time}
\end{figure}

\subsection{2-D HIT flame\label{subsec:subsection3}} \addvspace{10pt}

The DNN model for Jet-A/air combustion is examined using the configuration of HIT flame ignition case. Figure~\ref{Jet_HIT_T_Vor} depicts the contours of the temporal temperature distribution and the vorticity of the flow field for stoichiometric mixture to qualitatively compare the predictions of flame front structures and the effect of turbulence-chemistry interactions. As seen, the flame front propagation and flame wrinkling behaviour predicted using the DNN model show close agreement with those using CVODE. Next, quantitative comparison is performed on the turbulent burning velocity and presented in Fig.~\ref{Jet_HIT_St_time}. It can be observed that the turbulent burning rates still exhibit satisfactory agreement with a maximum error prediction of 7$\%$ for this more complex chemistry. 



\subsection{Computational acceleration analysis\label{subsec:subsection4}} \addvspace{10pt}

Following the above accuracy validations, the efficiency gain of the DNN approach is discussed. The overall computational time for a representative time period ( $t = $ 0.3 to 0.4 ms) in the propagating HIT flames is considered as an example. For the comparison shown here, the simulations with DI are run with 16 CPUs (AMD Zen1), while the simulations with DNN models use one GPU (NVIDIA RTX 4090) for inference. As seen in Table~\ref{HIT-acceleration}, the DNN models achieve a speed-up factor of approximately 50 on chemistry calculations and 11 on overall calculations using one GPU for the ethylene/air case. Additionally, a higher speed-up factor of 90 on chemistry calculations and 11.6 on overall is observed for the larger Jet-A/air mechanism. The percentage of chemistry in the overall time cost is significantly reduced from 87\% to 12\% when the DNN is employed. This highlights a superior computational acceleration with the DNN models and suggests an increasing speed-up effect with the complexity of the chemical mechanism.

\begin{table}[ht] \footnotesize
\renewcommand{\arraystretch}{1.2}
\centering
\caption{Computational time (hours) and speed-up factor of chemistry and total calculations for the 2-D HIT flames.}
\vspace{6 pt}
\begin{tabular}[t]{lcccc}
\hline
         &  \multicolumn{2}{c}{\text{C$_2$H$_4$}/air}&\multicolumn{2}{c}{\text{JetA}/air}\\
 &Chemistry&Total&Chemistry&Total\\ \hline
 CVODE& 4.6&5.3
&9&10.3\\
 DNN& 0.09& 0.48& 0.10&0.89\\
         Speed-up&  $50\times$&   $11\times$&$90\times$& $11.6\times$\\ \hline
\end{tabular}
\label{HIT-acceleration}
\end{table}

\section{Conclusions\label{sec:section4}} \addvspace{10pt}

This work proposed a consistent and robust ML methodology designed for developing DNN models across diverse fuels and turbulent flames, and comprehensive validations were conducted to evaluate the model precision and generalisation capabilities. The methodology involves collecting thermochemical base states from a small set of canonical laminar flames. To enhance the coverage of these base states in composition space for high-dimensional flames, an effective strategy involving random perturbation and data augmentation is employed, along with an essential constraint on heat release rate changes to eliminate nonphysical perturbed states. The resulting high-quality training samples serve as the input in the DNN training, wherein special considerations of physical principles, including mass fraction unity conservation, enthalpy conservation, and error constraints on heat release rate, are integrated into the hyperparameter optimisation process. A thorough validation process was conducted, encompassing a range of laminar and turbulent flame configurations, across different chemical systems from simple hydrogen to complex jet fuel. The results obtained from the DNN models generally exhibit excellent agreement with those from direct integration, showcasing the high accuracy and generalisation of the proposed deep learning approach as a versatile tool for chemical reaction rate integration. Furthermore, computational acceleration using this approach provides a promising speed-up factor of 50 to 90 for chemistry integration and approximately 10 for the overall simulation. This observation indicates that stiff chemistry calculations will no longer be the computational bottleneck for turbulent flame simulations with high-precision DNN models.

While the approach and results presented in this work seem quite encouraging, there remains room to further improve the DNN model accuracy, particularly for complex fuels. Future work will also be directed towards model compression and optimisation to further increase the computational efficiency. Additionally, ongoing efforts are being dedicated to incorporating model inference into the solving procedure of partial differential equations (PDEs), aiming to achieve a full computation on the GPU. With the memory copy overhead eliminated, a further one or two orders of magnitude speed-up can be achieved. 

\acknowledgement{Declaration of competing interest} \addvspace{10pt}

The authors declare that they have no known competing financial interests or personal relationships that could have appeared to influence the work reported in this paper.



\acknowledgement{Supplementary material} \addvspace{10pt}

A supplementary file is attached.


 \footnotesize
 \baselineskip 9pt


\bibliographystyle{pci}
\bibliography{PCI_LaTeX}

\begin{thebibliography}{10}
\expandafter\ifx\csname url\endcsname\relax
  \def\url#1{\texttt{#1}}\fi
\expandafter\ifx\csname urlprefix\endcsname\relax\def\urlprefix{URL }\fi
\expandafter\ifx\csname href\endcsname\relax
  \def\href#1#2{#2} \def\path#1{#1}\fi

\bibitem{poinsot2005theoretical}
T.~Poinsot, D.~Veynante, Theoretical and numerical combustion, RT Edwards, Inc., 2005.

\bibitem{lam1994csp}
S.~Lam, D.~Goussis, The csp method for simplifying kinetics, Int. J. Chem. Kinet. 26~(4) (1994) 461--486.

\bibitem{Sun10}
W.~Sun, Z.~Chen, X.~Gou, Y.~Ju, A path flux analysis method for the reduction of detailed chemical kinetic mechanisms, Combust. Flame 157~(7) (2010) 1298--1307.

\bibitem{Lu05}
T.~Lu, C.~Law, A directed relation graph method for mechanism reduction, Proc. Combust. Inst. 30~(1) (2005) 1333--1341.

\bibitem{Pepiot08}
P.~Pepiot-Desjardins, H.~Pitsch, An efficient error-propagation-based reduction method for large chemical kinetic mechanisms, Combust. Flame 154~(1-2) (2008) 67--81.

\bibitem{ihme2022combustion}
M.~Ihme, W.~T. Chung, A.~A. Mishra, Combustion machine learning: Principles, progress and prospects, Prog. Energy Combust. Sci. 91 (2022) 101010.

\bibitem{christo1995utilising}
F.~Christo, A.~Masri, E.~Nebot, T.~Tur{\'a}nyi, Utilising artificial neural network and repro-modelling in turbulent combustion, in: Proceedings of ICNN'95-International Conference on Neural Networks, Vol.~2, IEEE, 1995, pp. 911--916.

\bibitem{blasco1998modelling}
J.~Blasco, N.~Fueyo, C.~Dopazo, J.~Ballester, Modelling the temporal evolution of a reduced combustion chemical system with an artificial neural network, Combust. Flame 113~(1-2) (1998) 38--52.

\bibitem{blasco2000self}
J.~A. Blasco, N.~Fueyo, C.~Dopazo, J.~Chen, A self-organizing-map approach to chemistry representation in combustion applications, Combust. Theory Model. 4~(1) (2000) 61.

\bibitem{sen2009turbulent}
B.~A. Sen, S.~Menon, Turbulent premixed flame modeling using artificial neural networks based chemical kinetics, Proc. Combust. Inst. 32~(1) (2009) 1605--1611.

\bibitem{sen2010linear}
B.~A. Sen, S.~Menon, Linear eddy mixing based tabulation and artificial neural networks for large eddy simulations of turbulent flames, Combust. Flame 157~(1) (2010) 62--74.

\bibitem{chatzopoulos2013chemistry}
A.~Chatzopoulos, S.~Rigopoulos, A chemistry tabulation approach via rate-controlled constrained equilibrium (rcce) and artificial neural networks (anns), with application to turbulent non-premixed ch4/h2/n2 flames, Proc. Combust. Inst. 34~(1) (2013) 1465--1473.

\bibitem{franke2017tabulation}
L.~L. Franke, A.~K. Chatzopoulos, S.~Rigopoulos, Tabulation of combustion chemistry via artificial neural networks (anns): Methodology and application to les-pdf simulation of sydney flame l, Combust. Flame 185 (2017) 245--260.

\bibitem{wan2020chemistry}
K.~Wan, C.~Barnaud, L.~Vervisch, P.~Domingo, Chemistry reduction using machine learning trained from non-premixed micro-mixing modeling: Application to dns of a syngas turbulent oxy-flame with side-wall effects, Combust. Flame 220 (2020) 119--129.

\bibitem{ding2021machine}
T.~Ding, T.~Readshaw, S.~Rigopoulos, W.~Jones, Machine learning tabulation of thermochemistry in turbulent combustion: An approach based on hybrid flamelet/random data and multiple multilayer perceptrons, Combust. Flame 231 (2021) 111493.

\bibitem{readshaw2021modeling}
T.~Readshaw, T.~Ding, S.~Rigopoulos, W.~Jones, Modeling of turbulent flames with the large eddy simulation--probability density function (les--pdf) approach, stochastic fields, and artificial neural networks, Phys. Fluids 33~(3) (2021).

\bibitem{zhang2022multi}
T.~Zhang, Y.~Yi, Y.~Xu, Z.~X. Chen, Y.~Zhang, E.~Weinan, Z.-Q.~J. Xu, A multi-scale sampling method for accurate and robust deep neural network to predict combustion chemical kinetics, Combust. Flame 245 (2022) 112319.

\bibitem{burke2012comprehensive}
M.~P. Burke, M.~Chaos, Y.~Ju, F.~L. Dryer, S.~J. Klippenstein, Comprehensive h2/o2 kinetic model for high-pressure combustion, Int. J. Chem. Kinet. 44~(7) (2012) 444--474.

\bibitem{wu2017develop}
K.~Wu, W.~Yao, X.~Fan, Development and fidelity evaluation of a skeletal ethylene mechanism under scramjet-relevant conditions, Energy Fuels 31~(12) (2017) 14296--14305.

\bibitem{xu2018physics}
R.~Xu, K.~Wang, S.~Banerjee, J.~Shao, T.~Parise, Y.~Zhu, S.~Wang, A.~Movaghar, D.~J. Lee, R.~Zhao, et~al., A physics-based approach to modeling real-fuel combustion chemistry--ii. reaction kinetic models of jet and rocket fuels, Combust. Flame 193 (2018) 520--537.

\bibitem{Mao2023DeepFlame}
R.~Mao, M.~Lin, Y.~Zhang, T.~Zhang, Z.-Q. Xu, Z.~Chen, Deepflame: A deep learning empowered open-source platform for reacting flow simulations, Comput. Phys. Commun. 291 (2023) 108842.

\bibitem{box1964analysis}
G.~E. Box, D.~R. Cox, An analysis of transformations, J. R. Stat. Soc. Ser. B Methodol. 26~(2) (1964) 211--243.

\bibitem{driscoll2020premixed}
J.~F. Driscoll, J.~H. Chen, A.~W. Skiba, C.~D. Carter, E.~R. Hawkes, H.~Wang, Premixed flames subjected to extreme turbulence: Some questions and recent answers, Prog. Energy Combust. Sci. 76 (2020) 100802.

\bibitem{saito2023data}
M.~Saito, J.~Xing, J.~Nagao, R.~Kurose, Data-driven simulation of ammonia combustion using neural ordinary differential equations (node), Appl. Energy Combust. Sci. 16 (2023) 100196.

\bibitem{Karimkashi2020a}
S.~Karimkashi, H.~Kahila, O.~Kaario, M.~Larmi, V.~Vuorinen, A numerical study on combustion mode characterization for locally stratified dual-fuel mixtures, Combust. Flame 214 (2020) 121--135.

\bibitem{Vuorinen2016DNSLab}
V.~Vuorinen, K.~Keskinen, Dnslab: A gateway to turbulent flow simulation in matlab, Comput. Phys. Commun. 203 (2016) 278--289.

\end{thebibliography}


\newpage

\small
\baselineskip 10pt



\end{document}